\documentclass[twocolumn,aps,superscriptaddress,showpacs]{revtex4}
\usepackage{amssymb}
\usepackage{amsmath}
\usepackage{graphicx}
\usepackage[normalem]{ulem}
\usepackage[dvips]{color}

\renewcommand\sout{\bgroup \color{red} \ULdepth=-.5ex \ULset}

\begin{document}

\title{Triton-$^3$He relative and differential flows as probes of the nuclear symmetry energy at supra-saturation densities}
\author{Gao-Chan Yong}
\affiliation{Department of Physics and Astronomy, Texas A\&M
University-Commerce, Commerce, TX 75429-3011, USA}
\affiliation{Institute of Modern Physics, Chinese Academy of
Sciences, Lanzhou 730000, China}
\author{Bao-An Li\footnote{Corresponding author: Bao-An\_Li@Tamu-Commerce.edu}}
\affiliation{Department of Physics and Astronomy, Texas A\&M
University-Commerce, Commerce, TX 75429-3011, USA}
\author{Lie-Wen Chen}
\affiliation{Department of Physics, Shanghai Jiao Tong University,
Shanghai 200240, China}
\affiliation{Center of Theoretical Nuclear
Physics, National Laboratory of Heavy Ion Accelerator, Lanzhou
730000, China}
\author{Xun-Chao Zhang}
\affiliation{Institute of Modern Physics, Chinese Academy of
Sciences, Lanzhou 730000, China}
\date{\today}

\begin{abstract}
Using a transport model coupled with a phase-space coalescence
after-burner we study the triton-$^3$He ratio, relative and
differential transverse flows in semi-central $^{132}Sn+^{124}Sn$
reactions at a beam energy of $400$ MeV/nucleon. The
neutron-proton ratio, relative and differential flows are also
discussed as a reference. We find that similar to the
neutron-proton pairs the triton-$^3$He pairs also carry
interesting information about the density dependence of the
nuclear symmetry energy. Moreover, the nuclear symmetry energy
affects more strongly the t-$^3$He relative and differential flows
than the $\pi^-/\pi^+$ ratio in the same reaction. The t-$^3$He
relative flow can be used as a particularly powerful probe of the
high-density behavior of the nuclear symmetry energy.
\end{abstract}

\maketitle

\section{Introduction}

The density dependence of nuclear symmetry energy especially at
supra-saturation densities is among the most uncertain properties
of neutron-rich nuclear matter\cite{Kut94,Kub99}. However, it is
very important for nuclear structure\cite{Bro00,Hor01}, heavy-ion
reactions\cite{LiBA98,LiBA01b,Dan02a,Bar05,CKLY07,LCK08} and many
phenomena/processes in astrophysics and cosmology
\cite{Sum94,Lat04,Ste05a}. Heavy-ion reactions especially those
induced by radioactive beams provide a unique opportunity to
constrain the symmetry energy at supra-saturation densities in
terrestrial laboratories. Various probes using heavy-ion reactions
have been proposed in the literature, see, e.g., ref.\cite{LCK08}
for the most recent review. It is particularly interesting to
mention that, besides many significant results about the symmetry
energy at sub-saturation densities, see, e.g.,
refs.\cite{chen04,li05,chen05,She07,Tsang09,Cen09,Leh09},
circumstantial evidence for a rather soft symmetry energy at
supra-saturation densities has been reported very
recently\cite{xiao09} based on the IBUU04 transport
model\cite{IBUU04} analysis of the $\pi^-/\pi^+$ data taken by the
FOPI Collaboration at SIS/GSI\cite{Rei07}. However, many
interesting issues remain to be resolved. Thus, to constrain
tightly and reliably the nuclear symmetry energy especially at
supra-saturation densities, much more efforts by both the nuclear
physics and the astrophysics communities are still needed.

In the present work, we have the following two main purposes.
Firstly, there is an urgent need to verify the conclusion about
the soft symmetry energy at supra-saturation densities required to
reproduce the FOPI $\pi^-/\pi^+$ data within transport model
analyses\cite{xiao09,Rei07}. It is better if this test can be done
with not only more $\pi^-/\pi^+$ data but also other sensitive
observables in the most neutron-rich reactions possible. We will
thus make predictions using the same transport model\cite{IBUU04}
for doing this test. Secondly, it was predicted that the
neutron-proton differential flow is another sensitive probe of the
high-density behavior of the nuclear symmetry energy\cite{ba00a}.
However, it is difficult to measure observables involving
neutrons. One question often asked by some experimentalists is
whether the triton-$^3$He pair may carry the same information as
the neutron-proton one. We will try to answer this question
quantitatively by coupling the IBUU04 calculations to a
phase-space coalescence after-burner. Indeed, we found that,
similar to the neutron-proton pair, the triton-$^{3}$He relative
and differential transverse flows are sensitive to the
high-density behavior of the nuclear symmetry energy. They can be
used to test indications about the high-density behavior of the
symmetry energy observed earlier from analyzing the $\pi^-/\pi^+$
data.

\section{Summary of theoretical models}
Our study is carried out based on the IBUU04 version of an isospin
and momentum dependent transport model and the simplest
phase-space coalescence after-burner. For completeness and
consistency we recall here a few major features of the IBUU04
transport model most relevant to the present study. More details
of the model can be found in Refs. \cite{IBUU04}. The single
nucleon potential is one of the most important inputs to BUU-like
transport models for nuclear reactions. In the IBUU04 transport
model, we use a single nucleon potential derived within the
Hartree-Fock approach using a modified Gogny effective interaction
(MDI) \cite{das03}, i.e.,
\begin{widetext}
\begin{eqnarray}\label{mdi}
U(\rho ,\delta ,\mathbf{p},\tau ) &=&A_{u}(x)\frac{\rho _{\tau ^{\prime }}}{%
\rho _{0}}+A_{l}(x)\frac{\rho _{\tau }}{\rho _{0}}
+B(\frac{\rho }{\rho _{0}})^{\sigma }(1-x\delta ^{2})-8x\tau \frac{B}{%
\sigma +1}\frac{\rho ^{\sigma -1}}{\rho _{0}^{\sigma }}\delta \rho
_{\tau^{\prime }}  \nonumber \\
&&+\frac{2C_{\tau ,\tau }}{\rho _{0}}\int d^{3}\mathbf{p}^{\prime }\frac{%
f_{\tau}(\mathbf{r},\mathbf{p}^{\prime})}{1+(\mathbf{p}-\mathbf{p}^{\prime
})^{2}/\Lambda ^{2}}+\frac{2C_{\tau ,\tau ^{\prime }}}{\rho _{0}}\int d^{3}\mathbf{p}^{\prime }%
\frac{f_{\tau ^{\prime }}(\mathbf{r},\mathbf{p}^{\prime })}{1+(\mathbf{p}-%
\mathbf{p}^{\prime })^{2}/\Lambda ^{2}}.
\end{eqnarray}
\end{widetext}
Here $\delta =(\rho _{n}-\rho _{p})/\rho $ is the
isospin asymmetry of the nuclear medium. In the above $\tau =1/2$
($-1/2$) for neutrons (protons) and $\tau \neq \tau ^{\prime }$;
$\sigma =4/3$; $f_{\tau }(\vec{r},\vec{p})$ is the phase space
distribution
function at coordinate $\vec{r}$ and momentum $%
\vec{p}$. The parameter $x$ was introduced to mimic predictions on
the density dependence of symmetry energy $E_{\text{sym}}(\rho )$
by microscopic and/or phenomenological many-body theories. The
parameters $A_{u}(x)$ and $A_{l}(x)$ depend on the $x$ parameter
according to
\begin{equation}
A_{u}(x)=-95.98-x\frac{2B}{\sigma
+1},~~~~A_{l}(x)=-120.57+x\frac{2B}{\sigma +1}.
\end{equation}
The coefficients in $A_{u}(x)$ and $A_{l}(x)$ and the parameters
$B,C_{\tau ,\tau },C_{\tau ,\tau ^{\prime }}$ and $\Lambda $ were
obtained by fitting the momentum-dependence of the $U(\rho ,\delta
,\vec{p},\tau ,x)$ to that predicted by the Gogny Hartree-Fock
and/or the Brueckner-Hartree-Fock (BHF) calculations, the
saturation properties of symmetric nuclear matter and the symmetry
energy of about $31.6$ MeV at normal nuclear matter density $\rho
_{0}=0.16$ fm$^{-3}$. The incompressibility $K_{0}$ of symmetric
nuclear matter at $\rho _{0}$ is set to be $211$ MeV consistent
with the latest conclusion from studying giant resonances
\cite{das03}.

The last two terms in Eq. (\ref{mdi}) contain the
momentum-dependence of the single-particle potential. The momentum
dependence of the symmetry potential stems from the different
interaction strength parameters $C_{\tau ,\tau ^{\prime }}$ and
$C_{\tau ,\tau }$ for a nucleon of isospin $\tau $ interacting,
respectively, with unlike and like nucleons in the background
fields. More specifically, we use $C_u\equiv C_{unlike}=-103.4$
MeV and $C_l\equiv C_{like}=-11.7 $ MeV. With these parameters,
the isoscalar potential estimated from
$(U_{neutron}+U_{proton})/2$ agrees reasonably well with
predictions from the variational many-body theory
\cite{wiringa,IBUU04}. At the normal nuclear matter density
$\rho_0$, it is consistent with the isoscalar nucleon optical
obtained from the nucleon-nucleus scattering experiments
\cite{IBUU04}. The strengthes of the corresponding isovector
(symmetry) potential can be estimated from $(U_n-U_p)/2\delta$. At
$\rho_0$, by design, the symmetry potential is independent of x
and is consistent with the Lane potential extracted from the
nucleon-nucleus scattering experiments and the (p,n) charge
exchange reactions \cite{LCK08}.

The corresponding MDI symmetry energy can be written as
\cite{Xu09}
\begin{widetext}
\begin{eqnarray}\label{esymmdi}
E_{sym}(\rho)&=& \frac{8 \pi}{9 m h^3 \rho} p^5_f + \frac{\rho}{4
\rho_0} [-24.59+4Bx/(\sigma +1)] - \frac{B x}{\sigma + 1}
\left(\frac{\rho}{\rho_0}\right)^\sigma \nonumber\\
&+& \frac{C_l}{9 \rho_0 \rho} \left(\frac{4 \pi}{h^3}\right)^2
\Lambda^2 \left[4 p^4_f - \Lambda^2 p^2_f \ln \frac{4 p^2_f +
\Lambda^2}{\Lambda^2}\right] + \frac{C_u}{9 \rho_0 \rho}
\left(\frac{4 \pi}{h^3}\right)^2 \Lambda^2 \left[4 p^4_f - p^2_f
(4 p^2_f + \Lambda^2) \ln \frac{4 p^2_f +
\Lambda^2}{\Lambda^2}\right],
\end{eqnarray}
\end{widetext}
where $p_f=\hbar(3\pi^2\frac{\rho}{2})^{1/3}$ is the Fermi momentum
for symmetric nuclear matter at density $\rho$.

The IBUU04 model can use either the experimental nucleon-nucleon
(NN) cross sections or the in-medium NN cross sections calculated
using an effective-mass scaling model consistent with the single
particle potential used \cite{li05}. In the present work the
in-medium NN cross sections are used. The isospin-dependent Pauli
blocking has been implemented by evolving and checking explicitly
neutron and proton phase-space distributions separately. The
coordinates of nucleons in the colliding nuclei are initialized
randomly according to the neutron/proton density profiles
predicted by the Skyrme-Hartree-Fock approach. The corresponding
Fermi momenta are calculated using the local Thomas-Fermi
approximation. The initial state generated in such way is rather
stable for several hundred fm/c without appreciable particle
emission in evolving a single nucleus with a momentum-independent
mean-field. However, as it is widely known, see, e.g., ref,
\cite{Pur09}, momentum dependent mean-field makes the initial
state less stable. For instance, in evolving a single $^{124}$Sn
nucleus with the IBUU04, at 40 fm/c (by which most of the nucleons
should have freezed-out in heavy-ion reactions at a beam energy of
400 MeV/nucleon) the average ratios of emitted/initial protons and
neutrons are 0.6/50 and 1.7/74, respectively, in calculations
using 600 test-particles per nucleon. The ratios go up to about
1.5/50 and 3.4/74, respectively, in calculations using 200
test-particles per nucleon. Our results presented in the following
are obtained using totally 10,000 events in each case by using 200
test-particles per nucleon in each run of the simulation.

Because most BUU-type transport models including the IBUU04 are
incapable of forming dynamically realistic nuclear fragments, some
types of after-burners, such as statistical and coalescence models,
are normally used as a remedy. This kind of hybrid models can be
used to study reasonably well, for instance, nuclear
multifragmentation, see, e.g.,
ref.\cite{kruse85,Ligross,hagel,Tan01}, collective flow of light
fragments, see e.g., \cite{koch90,chen98,zhang99} and the formation
of hypernuclei\cite{gait08}. There are, however, some remaining
issues, such as the freeze-out time of fragments that is related to
the time of coupling the transport model with the after-burner, etc.
There are also interesting work in using advanced coalescence
models\cite{Mat97,Ru99}, see, e.g., refs.\cite{chen03,Chen04b}. We
notice here that, several advanced cluster recognition routines,
such as, the Early Cluster Recognition Algorithm (ECRA)
\cite{Str97}, the Simulated Annealing Clusterization Algorithm
(SACA) \cite{Pur00}, have been put forward in recent years. For the
purposes of the present exploration, however, we use the simplest
phase-space coalescence model, see, e.g.,
refs.\cite{chen98,zhang99}, where a physical fragment is formed as a
cluster of nucleons with relative momenta smaller than $P_{0}$ and
relative distances smaller than $R_{0}$. The results presented in
the following are obtained with $P_{0}=263$ MeV/c and $R_{0}=3$ fm.
This simple choice may thus limit the scope and importance of our
study here. For instance, we shall limit ourselves to studies of the
relative/differential observables for neutron-proton and t-$^3$He
pairs without attempting to study pairs of the heavier mirror
nuclei. An extended study including the heavy mirror nuclei using
the advanced coalescence and/or earlier cluster recognition methods
is planned.

\section{Results and discussions}
\begin{figure}[t]
\begin{center}
\includegraphics[width=0.5\textwidth]{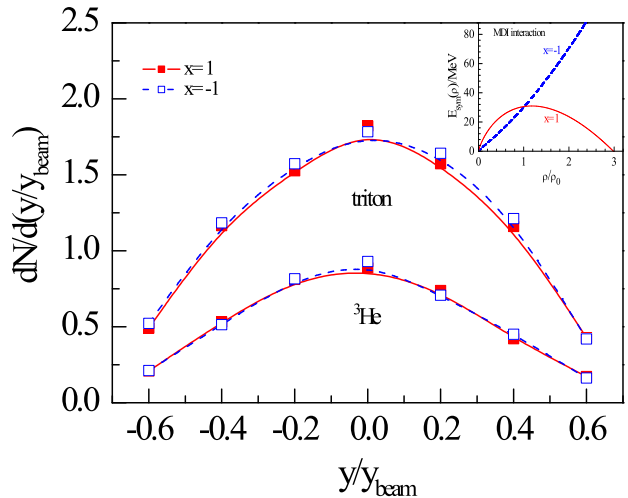}
\end{center}
\caption{Multiplicities of triton and $^{3}$He as a function of
the reduced C.M. rapidity in the reaction of $^{132}Sn+^{124}Sn$
at a beam energy of 400 MeV/nucleon and an impact parameter of 5
fm with the symmetry energy parameter $x= 1$ (soft) and $x= -1$
(stiff), respectively.} \label{multi}
\end{figure}

Noticing that the FOPI  $\pi^-/\pi^+$ data favors the symmetry
energy with $x=1$ at supra-saturation densities\cite{xiao09} while
the NSCL/MSU isospin diffusion data favors a symmetry energy at
sub-saturation densities between those with $x=0$ and
$x=-1$\cite{chen04,li05}, for comparisons we use here $x=1$ and
$x=-1$ as two limits of the symmetry energy at high densities. The
corresponding symmetry energy functionals are depicted in the
inset of Fig.\ \ref{multi}. They represent a typically stiff
($x=-1$) and a very soft ($x=1$) symmetry energy at
supra-saturation densities.

In the FOPI experiments\cite{Rei07}, the $\pi^-/\pi^+$ ratio was
measured down to a beam energy of 400 MeV/nucleon where it shows
the largest sensitivity to the high-density behavior of the
nuclear symmetry energy\cite{xiao09}. The maximum density reached
in central Au+Au reactions at 400 MeV/nucleon is about
$2.5\rho_0$. It is well known that pions are most abundantly
produced in the central collisions. However, the $\pi^-/\pi^+$
ratio is almost a constant from most central to mid-central impact
parameters\cite{xiao09}. It is also well known that transverse
collective flow is zero in head-on and grazing collisions but is
the largest in mid-central collisions. Considering all of the
above, as an example, we study $^{132}Sn+^{124}Sn$ reaction at an
incident beam energy of 400 MeV/nucleon and an impact parameter of
5 fm. This reaction will be available at FAIR/GSI in the near
future. We note that the maximum density reached in this reaction
is about $2\rho_0$.

Before we study the triton-$^3$He relative and differential flows,
it is necessary to first examine the yields of triton and $^3$He
and their ratio. Although these observables and their dependence
on the symmetry energy have been studied before especially at
lower incident energies, see, e.g., ref.\cite{chen03,Chen04b},
they serve as useful references for measuring the symmetry energy
effects on the $\pi^-/\pi^+$ ratio and the flow observables. Shown
in Fig.\ \ref{multi} are the rapidity distributions of triton and
$^{3}$He for the $^{132}Sn+^{124}Sn$ reaction at an incident
energy of 400 MeV/nucleon and an impact parameter of 5 fm. It is
seen that these clusters are produced mostly at mid-rapidities
from the participant region. The yields, however, are not
sensitive to the symmetry energy. This is not surprising. Even at
much lower energies where effects of the symmetry energy is
stronger, the neutron and proton yields themselves are not so
sensitive to the symmetry energy as the yields are dominated by
the isoscalar part of the nuclear mean field and nucleon-nucleon
collisions.

\begin{figure}[t]
\begin{center}
\includegraphics[width=0.5\textwidth]{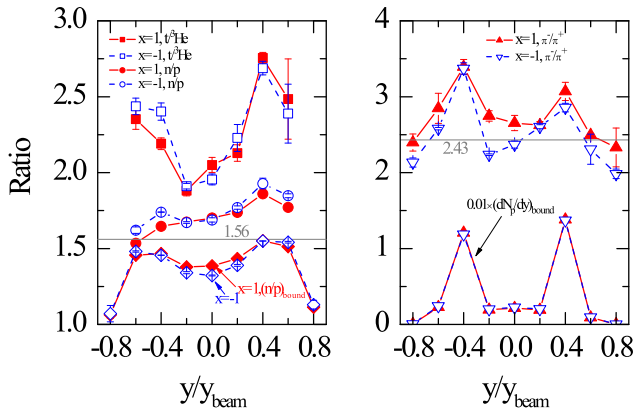}
\end{center}
\caption{Comparisons of particle ratios in the same reaction as in
Fig.\ \ref{multi}. The horizonal line at 1.56 (2.43) in the left
(right) window is the reaction system's neutron/proton ratio
(squared).} \label{rmulti}
\end{figure}
At Fermi energies, the ratio of neutrons to protons or that of
mirror nuclei were shown to carry more information about the
symmetry energy as effects of the isoscalar potential can be largely
cancelled in the ratios\cite{LiBA97a,chen03,Chen04b,IBUU04,Riz05}.
However, at significantly higher energies, e.g., 400 MeV/nucleon,
except for high transverse momentum particles especially those being
squeezed-out perpendicular to the reaction plane\cite{Yong67}, the
neutron to proton ratio becomes less sensitive to the symmetry
energy than at lower beam energies\cite{ba00a,Li06}. For
comparisons, the t$/^3$He ratio together with the free and bound
neutron to proton rations are shown in the left window of Fig.\
\ref{rmulti} for the $^{132}Sn+^{124}Sn$ reaction. Here, the bound
neutrons and protons are the nucleons in the fragments with $A\geq
2$. The horizonal line at 1.56 is the reaction system's
neutron/proton ratio $\mathrm{(N_T+N_P)/(Z_T+Z_P)}$. Several
interesting observations can be made here. Firstly, the
$(n/p)_{\mathrm{bound}}$ ($(n/p)_{\mathrm{free}}$) is significantly
less (higher) than the neutron/proton ratio of the reaction system.
This is the well known isospin fractionation
phenomenon\cite{Mul95,LiBA97b,Bar98} which is reduced here by the
production of charged pions. Moreover, at mid-rapidity the
$(n/p)_{\mathrm{bound}}$ shows appreciable sensitivity to the
variation of the symmetry energy. Unfortunately, both the
$(n/p)_{\mathrm{free}}$ and t/$^{3}$He ratios show very little
sensitivity to the variation of the symmetry energy within
statistical error bars, except around the projectile and target
rapidities of $y/y_{\mathrm{beam}}=\pm 0.5$.

It is especially worth noting that the t$/^3$He ratio is much
higher than the free and bound neutron/proton ratios. A few more
comments about this observation are in order here. First of all,
we notice that the assumption of $t/^3$He=$(n/p)_{\mathrm{free}}$
and that they are equal to the $\mathrm{(N_T+N_P)/(Z_T+Z_P)}$ have
been widely used in momentum-space coalescence models in the
literature, especially at high energies. Our results here and also
those reported earlier from transport model
simulations\cite{hagel,lee} indicate that the coordinate-space
correlation is important in considering the cluster formation.
This feature has also been observed in the parton hadronization
process in ultra-relativistic heavy-ion collisions by the quark
coalescence model\cite{chen06}. In fact, a similarly high $t/^3$He
ratio has also been observed in the experimental data of heavy-ion
reactions at Fermi energies, see, e.g., ref.\cite{hagel}. The
possible explanations for the high n/p and t$/^3$He ratios include
the isospin fractionation and the overwhelmingly preferential
production of symmetric light clusters such as deuteron and alpha
particles\cite{Ran} in these reactions. Our analyses by turning
on/off the coalescence after-burner, the Coulomb and symmetry
potentials indicate that both mechanism are at work. It is also
interesting to mentioning that, at Fermi energies, using a
freeze-out temperature extracted from the experiments the rather
high t$/^3$He ratio can also be well reproduced within a
statistical model\cite{Ber} assuming that the difference in the
chemical potentials of triton and $^3$He is dominated by their
Coulomb potentials\cite{hagel}.

The rapidity distributions of the $\pi^-/\pi^+$ ratio and the
bound protons are shown in the right window of Fig.\ \ref{rmulti}.
Interestingly, comparing the ratios of all particle-pairs shown in
both windows of Fig.\ \ref{rmulti}, it is obvious that the
$\pi^-/\pi^+$ ratio shows the highest sensitivity to the symmetry
energy. More quantitatively, effects of the symmetry energy on the
ratio of the total yields of charged pions is about 10\% by
varying the $x$ parameter from $-1$ to $1$. This is less than the
approximately 20\% effect observed in the head-on collisions
between two Au nuclei at the same beam energy. This is probably
because of the significantly smaller size in $^{132}Sn+^{124}Sn$
although it is more neutron-rich\cite{Zha09}. Moreover, by
comparing the rapidity distributions of the $\pi^-/\pi^+$ ratio
and the bound protons one sees clearly the well-known Coulomb
focusing effects on the $\pi^-/\pi^+$ ratio, namely, more
$\pi^-$'s ($\pi^+$'s) are attracted (repelled) towards (away) from
the target and projectile residues\cite{Sto86,Yon06} where most of
the protons are located in the semi-central reactions considered.

\begin{figure}[t]
\begin{center}
\includegraphics[width=0.5\textwidth]{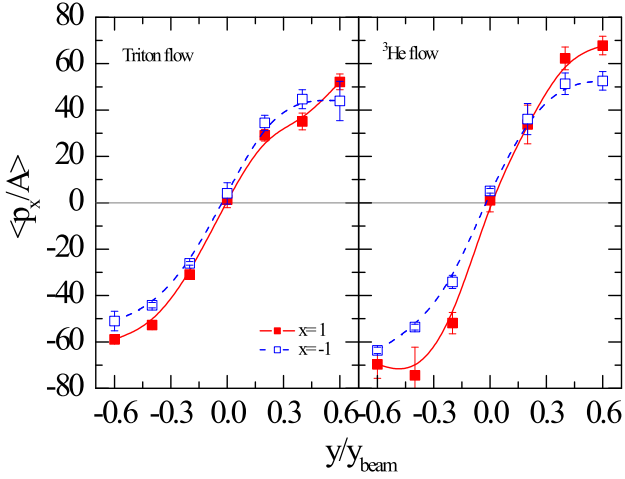}
\end{center}
\caption{Triton and $^{3}$He transverse flows (in unit of MeV) as
functions of the reduced C.M. rapidity in the same reaction as in
Fig.\ \ref{multi}.} \label{resflow}
\end{figure}
We now investigate whether the transverse collective flows of
triton and $^{3}$He can be used to probe the symmetry energy.
Firstly, we examine in Fig.\ \ref{resflow} their transverse flows
individually. The average C.M. transverse momentum per nucleon
$<p_{x}/A>$  in the reaction plane is defined as
\begin{eqnarray}
<p_{x}/A>(y) &\equiv &\frac{1}{N(y)}\sum_{i=1}^{N(y)}p_{x}^{i}/A(y)
\end{eqnarray}
where $N(y)$ is the total number of fragments of mass A in the
rapidity bin at $y$. The correlation between the $<p_{x}/A>$ and
rapidity $y$ reveals the transverse collective flow\cite{Dan85}.
It is seen that $^{3}$He clusters show a stronger flow than triton
clusters. This is mainly due to the stronger Coulomb force
experienced by the $^{3}$He clusters. More interestingly, the
transverse flow of $^{3}$He clusters show appreciable sensitivity
to the variation of the symmetry energy.

\begin{figure}[t]
\begin{center}
\includegraphics[width=0.5\textwidth]{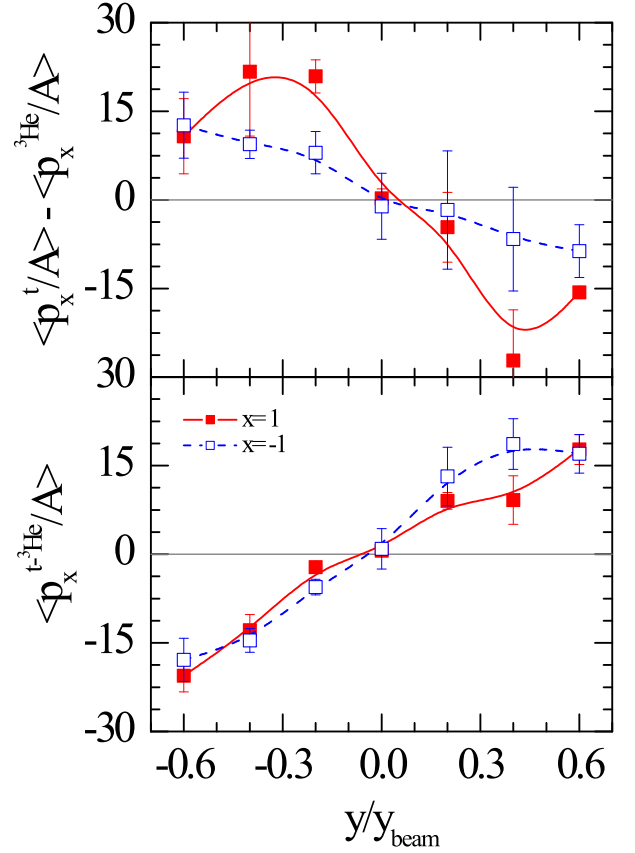}
\end{center}
\caption{Triton-$^{3}$He relative and differential flows (in unit
of MeV) as a function of the reduced C.M. rapidity in the same
reaction as in Fig.\ \ref{multi}.} \label{clusterflow}
\end{figure}

The transverse flow is a result of actions of several factors
including the isoscalar, symmetry and Coulomb potentials and
nucleon-nucleon scatterings. It is well known that the transverse
flow is sensitive to the isoscalar potential. Given the remaining
uncertainties associated with the isoscalar potential and the
small size of the symmetry energy effects, it would be very
difficult to extract any reliable information about the symmetry
energy from the individual flows of triton and $^{3}$He clusters.
Thus techniques of reducing effects of the isoscalar potential
while enhancing effects of the isovector potential are very
helpful \cite{ba00a,Yong67,Li06,gre03,fami06}. We thus study in
Fig.\ \ref{clusterflow} the triton-$^{3}$He relative and
differential flows. The relative flow is given as
\begin{eqnarray}
<p_{x}^{t}/A>-<p_{x}^{^{3}\mathrm{He}}/A>=
\frac{1}{N_{t}}\sum_{i=1}^{N_{t}}p_{x}^{i}/A-
\frac{1}{N_{^{3}\mathrm{He}}}\sum_{i=1}^{N_{^{3}\mathrm{He}}}p_{x}^{i}/A.
\end{eqnarray}
The triton-$^{3}$He differential flow  reads
\begin{eqnarray}
&&<p_{x}^{t-^{3}\mathrm{He}}/A>=
\frac{1}{N_{t}+N_{^{3}\mathrm{He}}}(\sum_{i=1}^{N_{t}}p_{x}^{i}/A-\sum_{i=1}^{N_{^{3}\mathrm{He}}}p_{x}^{i}/A)\nonumber\\
&=&\frac{N_{t}}{N_{t}+N_{^{3}\mathrm{He}}}<p_{x}^{t}/A>
-\frac{N_{^{3}\mathrm{He}}}{N_{t}+N_{^{3}\mathrm{He}}}<p_{x}^{^{3}\mathrm{He}}/A>,
\label{diflow}
\end{eqnarray}
where $N_{t}$, $N_{^{3}\mathrm{He}}$ are the number of triton and
$^{3}$He in the rapidity bin at $y$. From the upper panel of Fig.\
\ref{clusterflow}, it is seen that the triton-$^{3}$He relative
flow is very sensitive to the symmetry energy. Because of the
larger slope of the $^{3}$He flow, the triton-$^{3}$He relative
flow shows a negative slope at mid-rapidity. Effects of the
symmetry energy on the differential flow shown in the lower panel,
however, is relatively small. Although the $^{3}$He flow is more
sensitive to the symmetry energy, the small number of $^{3}$He
clusters (as shown in Fig.\ \ref{multi}) makes the $^{3}$He flow
contributes less to the triton-$^{3}$He differential flow (as
indicated in Eq. (\ref{diflow})). The triton-$^{3}$He differential
flow is therefore dominated by triton clusters. Consequently, it
is less sensitive to the symmetry energy than the triton-$^{3}$He
relative flow. The slope $F(x)\equiv d<p_x/A>/d(y/y_{beam})$ of
the transverse flow at mid-rapidity can be used to characterize
more quantitative the symmetry energy effects. We found that for
the t-$^3$He relative flow, $F(x=1)\approx-74$ MeV/c and
$F(x=-1)\approx-22$ MeV/c, respectively. For the t-$^3$He
differential flow, $F(x=1)\approx21$ MeV/c and $F(x=-1)\approx42$
MeV/c, respectively. Compared to the $\pi^-/\pi^+$ ratio in the
same reaction, the symmetry energy effects on the t-$^3$He
relative and differential flows are much stronger. Thus,
especially the t-$^3$He relative flow can be used as a very useful
and independent tool to test the soft symmetry energy at
supra-saturation densities extracted from studying the
$\pi^-/\pi^+$ ratio\cite{xiao09}.

\begin{figure}[t]
\begin{center}
\includegraphics[width=0.5\textwidth]{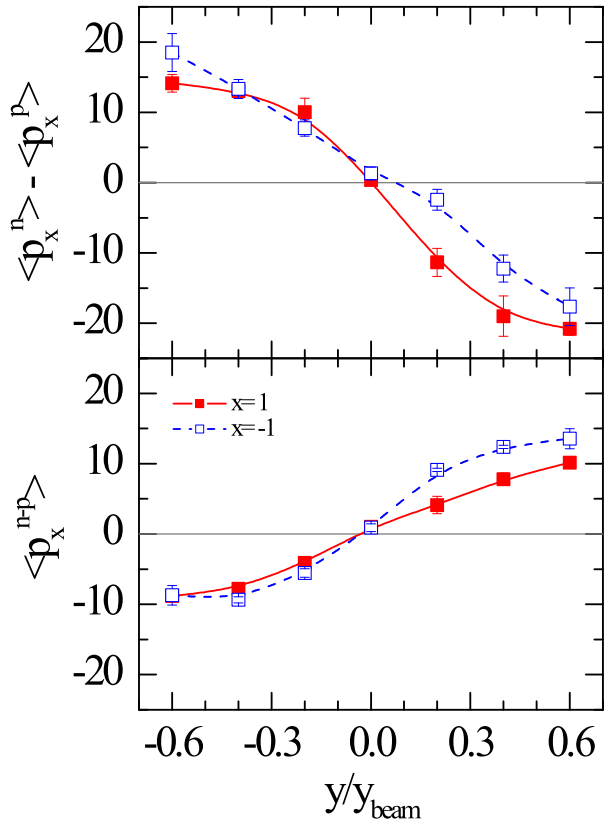}
\end{center}
\caption{Neutron-proton relative and differential flows (in unit
of MeV) as a function of the reduced C.M. rapidity in the same
reaction as in Fig.\ \ref{multi}.} \label{nucleonflow}
\end{figure}
For comparisons, we now study the relative and differential flows
for free neutron-proton pairs in Fig.\ \ref{nucleonflow}. It is
seen that they have the same features as the relative and
differential flows for triton-$^3$He pairs.  The larger symmetry
energy effects at positive rapidities are due to the more
neutron-rich of projectile. More quantitatively, in terms of the
slope parameter $F$, for the n-p relative flow, $F(x=1)\approx-53$
MeV/c and $F(x=-1)\approx-25$ MeV/c, respectively. For the n-p
differential flow, $F(x=1)\approx20$ MeV/c and $F(x=-1)\approx36$
MeV/c, respectively. Comparing the results in Figs.\
\ref{nucleonflow} and \ref{clusterflow} and the slope parameters
$F$ for t-$^3$He and neutron-proton pairs, we can conclude that
they are almost equally useful for probing the density dependence
of the nuclear symmetry energy.

\section{Summary}

In summary, using a hybrid approach coupling the transport model
IBUU04 to a phase-space coalescence after-burner we studied the
t-$^3$He relative and differential flows in semi-central
$^{132}Sn+^{124}Sn$ reactions at an incident energy of $400$
MeV/nucleon. We found that the nuclear symmetry energy affects
more strongly the t-$^3$He relative and differential flows than
the $\pi^-/\pi^+$ ratio in the same reaction. The t-$^3$He
relative flow can be used as a particular powerful probe of the
high-density behavior of the nuclear symmetry energy. It can be
used to test the indications about the symmetry energy at
supra-saturation densities observed in the analysis of the
$\pi^-/\pi^+$ data from heavy-ion reactions.

\begin{acknowledgments}
This work was supported in part by the US National Science
Foundation Awards PHY-0652548 and PHY-0757839, the Research
Corporation under Award No.7123 and the Texas Coordinating Board of
Higher Education Award No.003565-0004-2007, the National Natural
Science Foundation of China under grants 10710172, 10575119,
10675082 and 10975097 and MOE of China under project NCET-05-0392,
Shanghai Rising-Star Program under Grant No.06QA14024, the SRF for
ROCS, SEM of China, and the National Basic Research Program of China
(973 Program) under Contract No.2007CB815004.
\end{acknowledgments}


\end{document}